\documentstyle[preprint,aps,epsf]{revtex}
\begin{document}
\draft
\title{New Magic Numbers and New Islands of Stability in Drip-Line Regions}
\author{L. Satpathy$^{*}$}
\address{Institute of Physics, Bhubaneswar - 751 005, India}
\maketitle
\begin{abstract}

The systematics of the local energies and the two-neutron separation energies
obtained in the mass predictions of infinite nuclear matter model of atomic 
nuclei show strong evidence of new neutron magic numbers 100,152,164 , 
new proton magic number 78 and new islands of 
stability around N=100, Z $\simeq$ 62; N=152,Z=78; and N=164,Z$\simeq$90 
in the drip line regions of nuclear chart,where the usual magic numbers are 
found to be no more valid.

\pacs{PACS number: 21.10.Dr, 21.10.Pc, 21.60.Cs}
\end{abstract}

Nuclear matter which constitutes about 99.9$\%$ of the universe exists in many 
conditions in nature characterised by density, neutron-proton ratio and 
temperature. Our knowledge of nuclear dynamics has been mostly gathered 
during the last half a century or so , from the study of nuclei in the 
valley of 
stability. Obviously this knowledge cannot be considered as complete. Its 
adequacy and self-sufficiency has to be tested in newer and newer domains, and 
possibly improved upon,for its evolution towards completeness. 
Breakthroughs [1]achieved 
in nuclear techniques in recent years in the field of precision mass 
measurements, and production of radioactive ion beams have opened up the 
possibilities of study of exotic nuclear species in the drip-line regions 
of the nuclear "terra incognita" . An era of renaissance in nuclear structure 
physics has commenced with many opportunities and challenges paving the ways 
to perfect our 
knowledge of nuclear dynamics. Out of many questions of momentous nature, the 
most important one is whether  the classical magic numbers seen in the valley 
of stability remain valid in the exotic drip-line regions or new magic numbers 
and new islands of stability are awaiting to be discovered in such 
normally  inaccessible 
nuclear terrain . The nuclei inhabiting such regions will be precariously 
balanced in regard to their stability by the disruptive dripping force and the
stabilising shell effect arising out of the possible magicity. Such nuclei are produced 
in r-process nucleosynthesis. There abundances play important role in 
stellar evolution[2]. If such islands  exist, 
they will undoubtedly provide crucial new testing grounds for nuclear dynamics.
These are new nuclear terrains surrounded by relatively short lived exotic 
species unlike the superheavy elements which are true islands in the midst of 
sea of instability. This will provide new instance of the triumph of the 
individualistic properties of the nucleus over its liquid like behaviour in an 
exotic ambience like drip-line regions. In this note we present strong evidence of the existence 
of, new  magic numbers and new islands of stability found in our mass 
predictions[3] in Infinite Nuclear Matter(INM) model.[4,5]

The INM model of the nucleus is based on a many-body theoretic foundation 
whose main elements are the  infinite nuclear matter at ground state and the 
generalised Hugenholtz-Van Hove(HVH) theorem[6]. A good account of this can be seen
in [5].  
For easy reading and completeness, a brief account of INM model is 
presented here.
In this model the ground state energy
$E^{F}(A, Z)$ of a nucleus $(A, N, Z)$ with assymetry $\beta$ is considered
equivalent to the energy $E^{s}$ of a perfect sphere made up
of infinite nuclear matter at ground atate with the same assymetry $\beta$ plus
the residual characteristics energy $\eta$ called the local energy. So,
\begin{equation}
E^{F} (A, Z) = E^{s}_{INM} (A, Z) + \eta (A, Z)
\end{equation}
with
\begin{equation}
E^{s}_{INM} (A, Z) = E (A, Z) + f (A, Z)
\end{equation}
where $f (A, Z)$ characterises the finite size effects given by 
\begin{eqnarray}
f (A, Z) &=& a^{I}_{s} {A^{2/3}} + a^{I}_{c} (Z^2 - 5{{(3/16\pi)}^{2/3}} 
{Z^{4/3}}){A^{-1/3}} \\ \nonumber
&&  
- \delta (A, Z)
\end{eqnarray}

The superscript $I$ referes to the INM characteristics of the surface and
Coulomb co-efficients
 ${a^{I}}_s$ and ${a^{I}}_c$ respectively,
and $\delta (A, Z)$
is the usual pairing term. Hereafter the superscript `$F$' denotes the
quantities corresponding to finite nuclei.

Eq. (1) now becomes 
\begin{equation}
E^{F} (A, Z) = E (A, Z) + f (A, Z) + \eta (A, Z)
\end{equation}

        The  mass formula given by Eq.(4) consists of three distinct parts: an 
infinite part $E (A, Z)$, a finite size part $f (A, Z)$ and local energy 
part $\eta (A, Z)$. We have to determine these three functions. The term
$E (A, Z)$ being the property of INM at the ground state, will satisfy the
generalized  HVH theorem[6] of many-body theory,
\begin{equation}
\frac{E}{A} = [ ( 1+\beta ) {{\epsilon}_n} + ( 1-\beta ) {{\epsilon}_p} ]
\end{equation}
where, ${{\epsilon}_n} = {(\frac{\partial E}{\partial N}) \mid }_Z$ and         
${{\epsilon}_p} = {(\frac{\partial E}{\partial Z}) \mid }_N$ are neutron and proton 
Fermi energies respectively. The solution of Eq.(5) is of the form,
\begin{equation}
E = - a^{I}_v A + a^{I}_{\beta} {{\beta}^2} A
\end{equation}
where, $a^{I}_v$ and $a^{I}_{\beta}$ are identified as the volume and 
assymetry co-efficients corresponding to INM.

Using Eqs.(4), (5), and (6) we arrive at three essential equations of the
model,

\begin{equation}
\frac{f}{A} - {\frac{N}{A}} {(\frac{\partial f}{\partial N}) \mid }_Z
 - {\frac{Z}{A}} {(\frac{\partial f}{\partial Z}) \mid }_N
= \frac{E^F}{A} - {\frac{1}{2}} [ ( 1+\beta ) {{\epsilon}^{F}_n} + ( 1-\beta ) {{\epsilon}^{F}_p} ],
\end{equation}
\begin{equation}
- a^{I}_v + a^{I}_{\beta} {\beta ^2} =
{\frac{1}{2}} [ ( 1+\beta ) {{\epsilon}^{F}_n} + ( 1-\beta ) {{\epsilon}^{F}_p} ]
 - {\frac{N}{A}} {(\frac{\partial f}{\partial N}) \mid }_Z
 - {\frac{Z}{A}} {(\frac{\partial f}{\partial Z}) \mid }_N,
\end{equation}
and
\begin{equation}
\frac{\eta (A, Z)}{A} =
  {\frac{1}{2}} [ ( 1+\beta ) {(\frac{\partial \eta}{\partial N}) \mid }_Z
+ {\frac{1}{2}} [ ( 1-\beta ) {(\frac{\partial \eta}{\partial Z}) \mid }_N
\end{equation}

Eq.(7) determines the finite size coefficients characterising $\it f$ 
through its
fit to the combination of the data given by the r.h.s. of the equation, 
consisting of 
total energy and proton and neutron separation energies. These values so
determined are used 
in the fit of the Eq.(8) to determine $a_{v}^I$ and $a_{\beta}^I$, the 
properties of INM. Thus the coefficients $a_{v}^I$ , $a_{\beta}^I$ , 
 $a_{c}^I$ and $a_{s}^I$ determined through theses two fits are called global 
parameters , being the charecteristics of INM and common to all nuclei. $\eta$ 
denotes the characteristic properties of the nucleus which comprises shell ,
deformation and diffuseness etc. and can be considered as its finger print.
Once $E$ and $\it f$ are known, the empirical values of $\eta$ can be determined using 
experimental binding energies in Eq.(4) . Using the empirical values of 
$\eta$ of all known nuclei, the $\eta$ of all unknown nuclei are determined by 
using Eq.(9) for extrapolation. The details of the calculation can be found 
in [3].

        Over the years , the success of this mass formula has been well demonstrated
[3-12].
Being built over a many-body theoretic foundation, it has been shown to be 
particularly  useful in the extraction of the saturation properties 
and incompressibility 
of nuclear matter from nuclear masses leading to nuclear equation of state
[5]. 
Recently it has been developed [3] to its 
full potential in the prediction of masses of 7208 nuclei with  r.m.s.
and mean deviations of 401 keV and 9 keV respectively. The unique feature of 
this prediction is that it shows shell quenching[10] for the N=82, 
126 shells in 
agreement with the results of the astrophysical studies[2] of the abundances of 
heavy elements. The systematics of the local energy $\eta$ determined in this 
prediction have been shown by Nayak[12] to exhibit vanishing of the 
shells in the drip-line region. All these successes are essentially due to 
the inherent long range extrapolation properties of
 $\eta$ demonstrated explicitely for fifteen steps in [3].
This mass formula is quite special due to its strong many-body
theoretic foundation and its use of three times the data normally used 
in other mass formulas. It uses the binding energies and additionally, 
the neutron and proton separation energies. Thus it uses 5652 data for 
1884 nuclei in the fitting procedure to determine its five
parameters. Therefore it is endowed with better predictive power with
potential for validity in the regions far from stability. This is
evident in its unique success in shell-quenching[10] for higher shells 
$N=82,126$ where other mass formulas fail.
 This has 
naturally  enhanced 
our confidence in it . Here we make a thorough search of the systematics 
of $\eta$ and two-neutron separation energies in our mass predictions 
to see if any exotic features 
like new magic numbers and/or new islands of stability are existing in the 
drip-line regions of nuclear landscape. 

        The local energy  $\eta$ is a relatively new entity in nuclear 
physics. It comprises all the characeristic features of a nucleus, which  
includes  predominantly shell effect and all other  local 
effects like deformation, diffuseness etc. and possibly unknown ones also 
by its very construction. We 
would like to point out that the shell effect/energy depends upon the mean field 
one assigns to a nucleus. Since the nature of the mean field varies from 
region to region, its uniqueness is not global. In case of $\eta$,  the INM 
sphere which constitutes the bulk part of the total energy is made up of 
the infinite nuclear matter for all nuclei giving rise to its uniquely 
defined status. 
Therefore it is expected that, if nuclear landscape contains any specific 
local feature, it should be reflected in $\eta$ systematics. 

        Before we present our results, the general features of $\eta$ distributions 
as function of neutron and proton number are to be discussed. As shown 
by Nayak[12] and also will be shown here aposteriori, 
the plot of the values of $\eta$(N,Z) as a function of N for a given Z 
called the $\eta$ isoline, shows a Gaussian structure with the peak 
lying  at a magic number. The Gaussian peak on either side is flanked 
by the U-shape distribution corresponding to the isotopes of well deformed 
nuclei lying in the mid-shell region. The Gaussian peak gradually widens as 
one moves away from the magic number and finally disappears showing the 
shell-quenching effect. The U shapes also flatten up showing monotonic 
variation which implies  uniform shell structure and disappearance of 
nuclear magicity. The  width of the Gaussian peak may be taken as a measure 
of the  degree of magicity/shell closure. Thus it has been shown by Nayak 
recently that $\eta$ carries strong signature of nuclear shell structure.

         Another physical quantity namely the two-neutron separation energy $S_{2n}$
is known[13] to carry the signature of shell closure in the valley of stability.
The  $S_{2n}$ isolines show characteristic sharp bending just above and below 
the magic neutron number. Away from the magic number, these 
lines show only 
monotonic variations. It is interesting that $S_{2n}$ shell closure bending 
and $\eta$ Gaussian peak occur at the same magic number. Since, away from the 
magic number $S_{2n}$ shows only monotonic variation, while $\eta$ distribution shows U shape structure which sensitively varies with N and Z   
, $\eta$ 
distribution has emerged as a prominent and decisive signature of shell 
structure. 

        Here we have made a thorough search of the $\eta$ and $S_{2n}$ systematics 
in our 
mass predictions and obtained evidence of new shell closures and new islands 
of stability.The results are presented in Figs.1-4 for even values of Z only
to avoid clumsiness. The isolines for the intervening odd-Z ones lie between
the neighbouring even-Z ones at appropriate places. It may be noted that the 
specific elements and the range of their isotopes chosen for presentation here 
enables one to see in one sweep how the magicity evolves as one moves from 
the valley of stability to the drip-line. In Fig.1, $S_{2n}$ 
isolines for all the elements with even charge numbers  Z=50 $\sim$ 62  
are plotted for neutron numbers N=60 $\sim$ 110. In Fig.2, $\eta$ 
distributions  are 
shown for the same isotopes. This domain includes the well known magic numbers N=82 
and Z=50, and their  neighbourhoods for which the behaviour of $\eta$ and 
$S_{2n}$ distributions can be seen to be in accord with the discussions above. 
The characteristic features of $\eta$ Gaussians peaking around the 
magic numbers 
Z=50 , N=82 and their gradual widening as one moves away is clearly evident. 
This peak is flanked on either side by U shape distribution. The sharp shell 
closure bending in $S_{2n}$ distribution at N=82 in Fig.1 clearly 
correlates with the peak in the $\eta$ distribution in Fig.2. It is interesting to find 
in Fig.2 that the $\eta$ distributions for Z= 58, 60 and 62 show clear 
Gaussian peaks around N= 100. The corresponding $S_{2n}$ isolines in Fig.1 
show shell-closure type bending for these three elements at the same neutron 
number 100 , though not that prominently as at N= 82, but neverthless quite 
conspicuously. Hence we take neutron number 100 as a shell closure. As for 
the proton, none of the three Gaussians are strongly peaked to qualify for a 
good shell closure, however they represent somewhat weak magicity which may 
be reminiscent of their  
proximity to Z= 64 shell closure seen in the valley of stability. Hence 
we identify a new island of stability around N= 100, Z$\simeq$ 62. It is 
interesting to observe in Fig.2 how the magicity of the elements Z= 50 
varies with neutron number starting from N= 60 to N=110 manifesting its 
strongest magicity for N= 82 . However in the drip-line regions it 
does not remain a valid magic number. This is the manifestation of shell 
quenching.    

        In Figs.3 and 4, the $S_{2n}$ and $\eta$ isolines are depicted for the 
domain Z=78 $\sim$ 90 and N=100 $\simeq$ 170. This domain includes the 
well known magic 
numbers Z=82 and N=126 around which clear Gaussian distribution in Fig.4,
and sharp bending in Fig.3 are seen as expected. It is indeed pleasing to 
find two more Gaussian distributions around N=152  and N=164. The $\eta$ 
isolines for the four elements Z= 84, 82, 80, 78 in Fig.4 show well defined 
Gaussian peaks with increasing sharpness and heights around N= 152 as one 
moves down from 84 to 78. It is interesting to note that the quality of peak 
for Z= 78 around N= 152 is as good and even better than the peak for  
Z= 82 around N= 126 in the valley of stability. The $S_{2n}$ isolines in 
Fig.3 for these four elements show shell-closure type bending at N= 152.
Hence without reservations we assign magic number to N= 152 and Z=78 and 
identify a new island of stability around them.

        It is interesting to find in Fig.4 a  Gaussian peak for the element 
Z=90 lying in the fringe of the drip region around N=164. Although it is of 
small height, it is quite conspicuous . We find the element Z=88, 89 
exhibiting  peak structure also  around the 
same neutron number(Z=99 case not shown in the Fig.). The $S_{2n}$ isolines 
in Fig.3 show clear shell-
closure type bending for Z= 90,  88 at N= 164. The neutron number 164 has been 
anticipated before to be a magic number on the basis of theoretical studies
[14,15].
The present study which can be considered as an empirical one supports it. 
We will consider 90 as a poor  proton magic number. Thus a new island of 
stability around   N= 164 and 
Z$\simeq$ 90 is evident . 

        It may be observed in Fig.4 how the the proton number Z= 82 exhibits 
strongest magicity for  neutron number N= 126 in the valley of stability, and 
looses this property in the drip-line region. On the otherhand, the element 
Z= 78 which does not show magicity in the valley of stability , emerges as a
strong magic number in the drip line region. Thus the classical magic numbers 
no longer retain their magicity in the drip-line regions, and are replaced by 
new magic numbers.  

        In conclusion the present study provides strong evidence for new neutron 
magic numbers 100, 152 and 164, proton magic number 78 and new islands  of 
stability around N= 100, Z$\simeq$ 62; N= 152, Z= 78, and N= 164, Z$\simeq$ 90 
in the drip-line regions. This has been possible due to the uniqueness of 
the local energy $\eta$ in the INM model and its long range extrapolation 
property. The classical magic numbers do not remain valid in the exotic 
drip-line regions where new magic numbers make their appearance. The study gives the empirical evidence of the prepoderance of the 
individualistic properties like shell , deformation and diffuseness  etc. 
in the drip-line regions governing
the nuclear property in a decissive manner.

        The author acknowledges the financial support from the Council of Scientific 
and Industrial Research, Government of India .

\newpage

\begin{figure}
\vspace*{2.0cm}
\begin{center}
\epsfbox{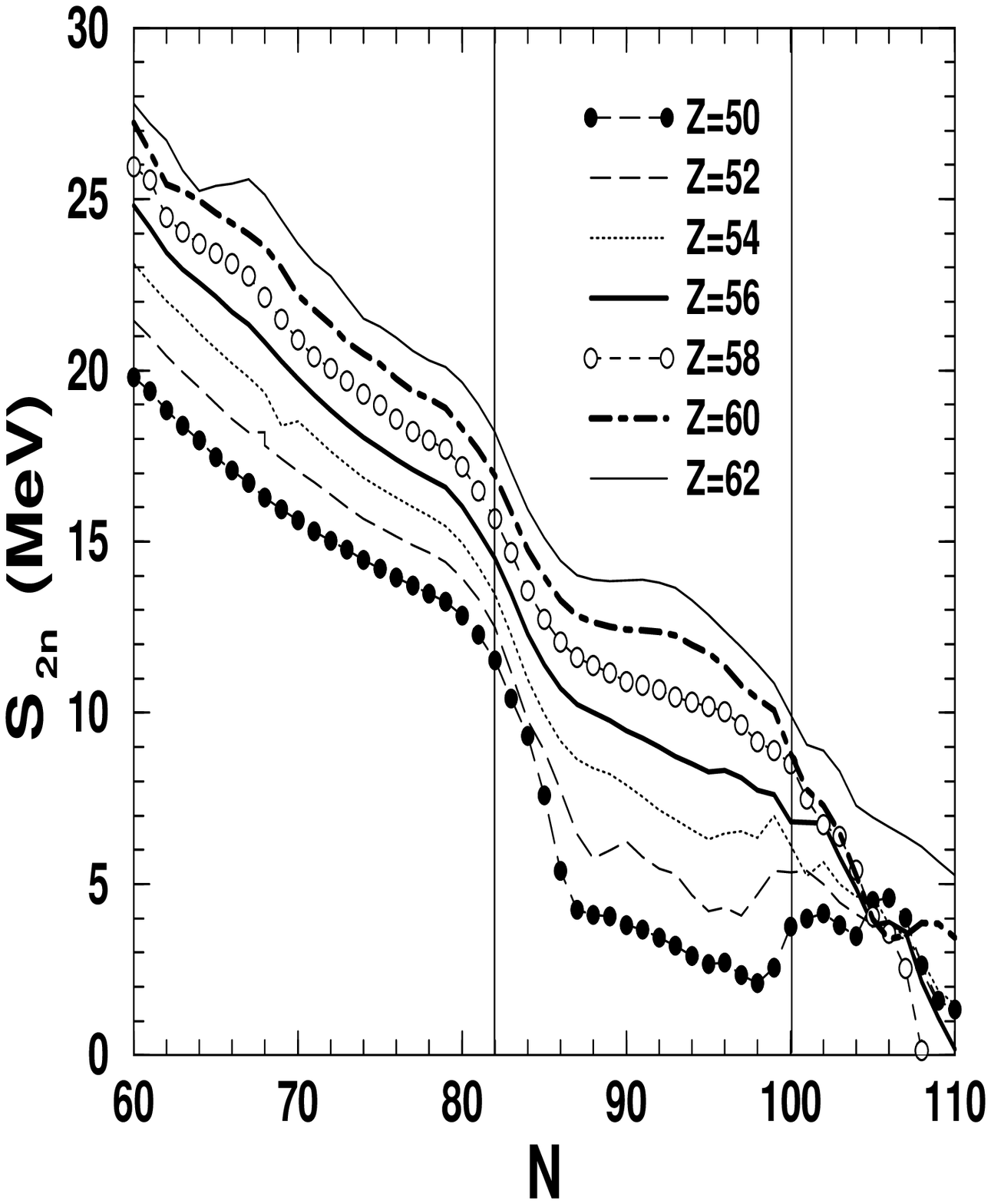}
\end{center}
\caption{The $S_{2n}$ isolines obtained in the INM model for $7$
  elements as function of neutron number $N$ for a series of
  isotopes. The vertical lines represent the neutron magic numbers $82$,
  and $100$.}
\label{1}
\end{figure}

\newpage

\vspace*{2.5cm}
\begin{figure}
\begin{center}
 \epsfbox{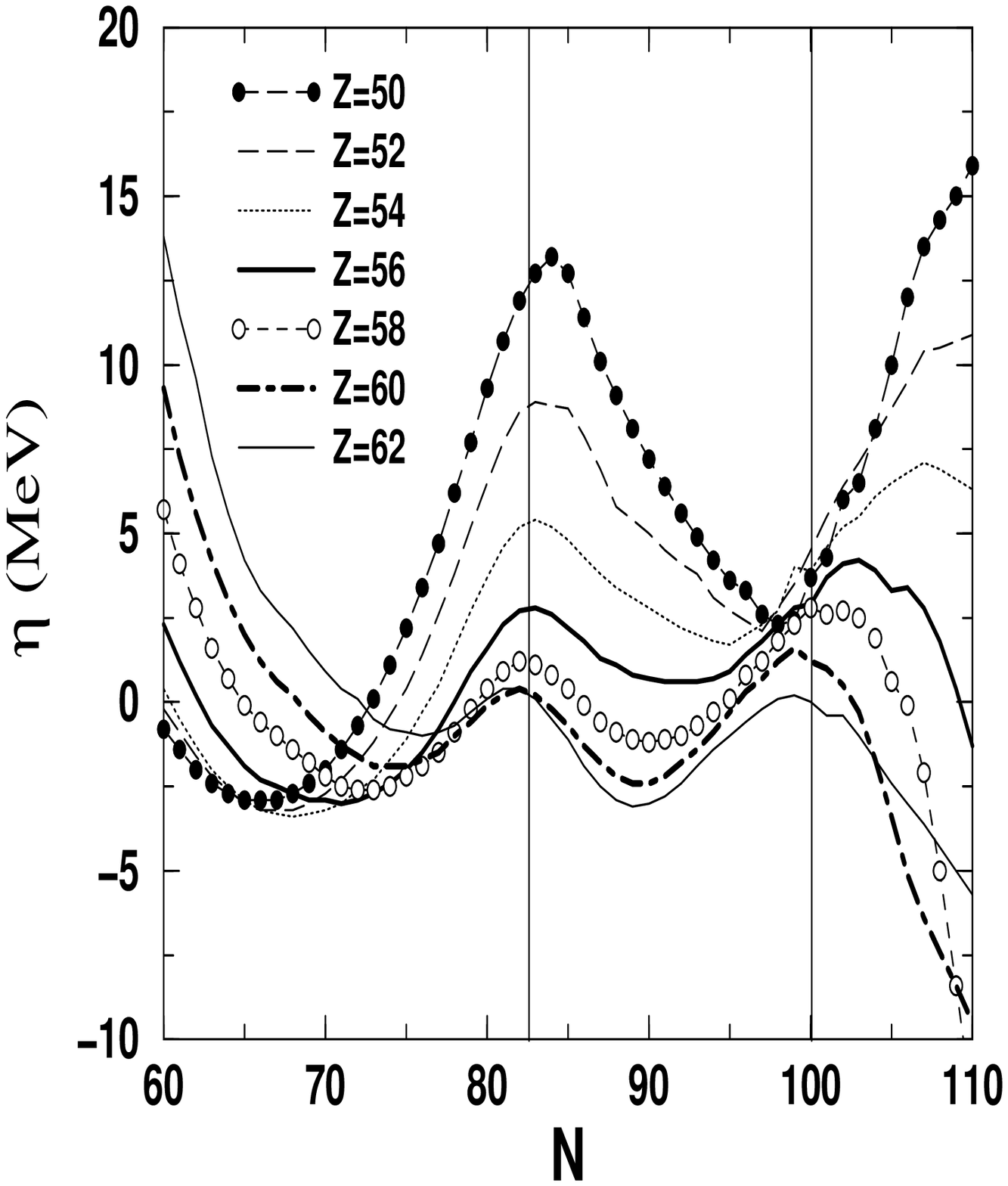}
\end{center}
\caption{Same as Fig 1 but for $\eta$ isolines. }
\label{2}
\end{figure}

\newpage

\vspace*{2.5cm}
\begin{figure}
\begin{center}
\epsfbox{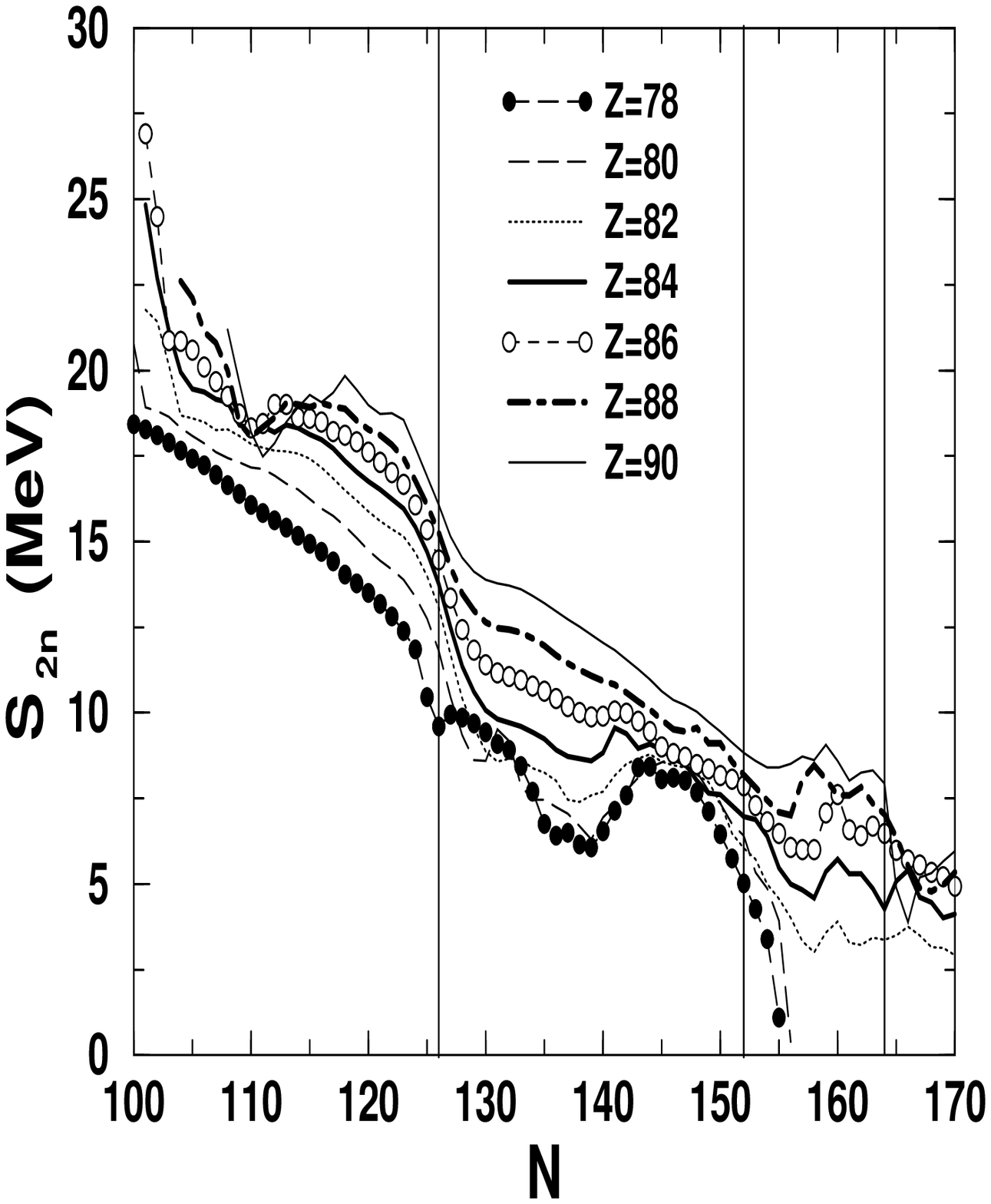}
\end{center}
\caption{The $S_{2n}$ isolines obtained in the INM model for $7$
  elements as function of neutron number $N$ for a series of
  isotopes. The vertical lines represent the neutron magic numbers $126$,
  $152$, and $164$.}
\label{3}
\end{figure}

\newpage

\vspace*{2.5cm}
\begin{figure}
\begin{center}
\epsfbox{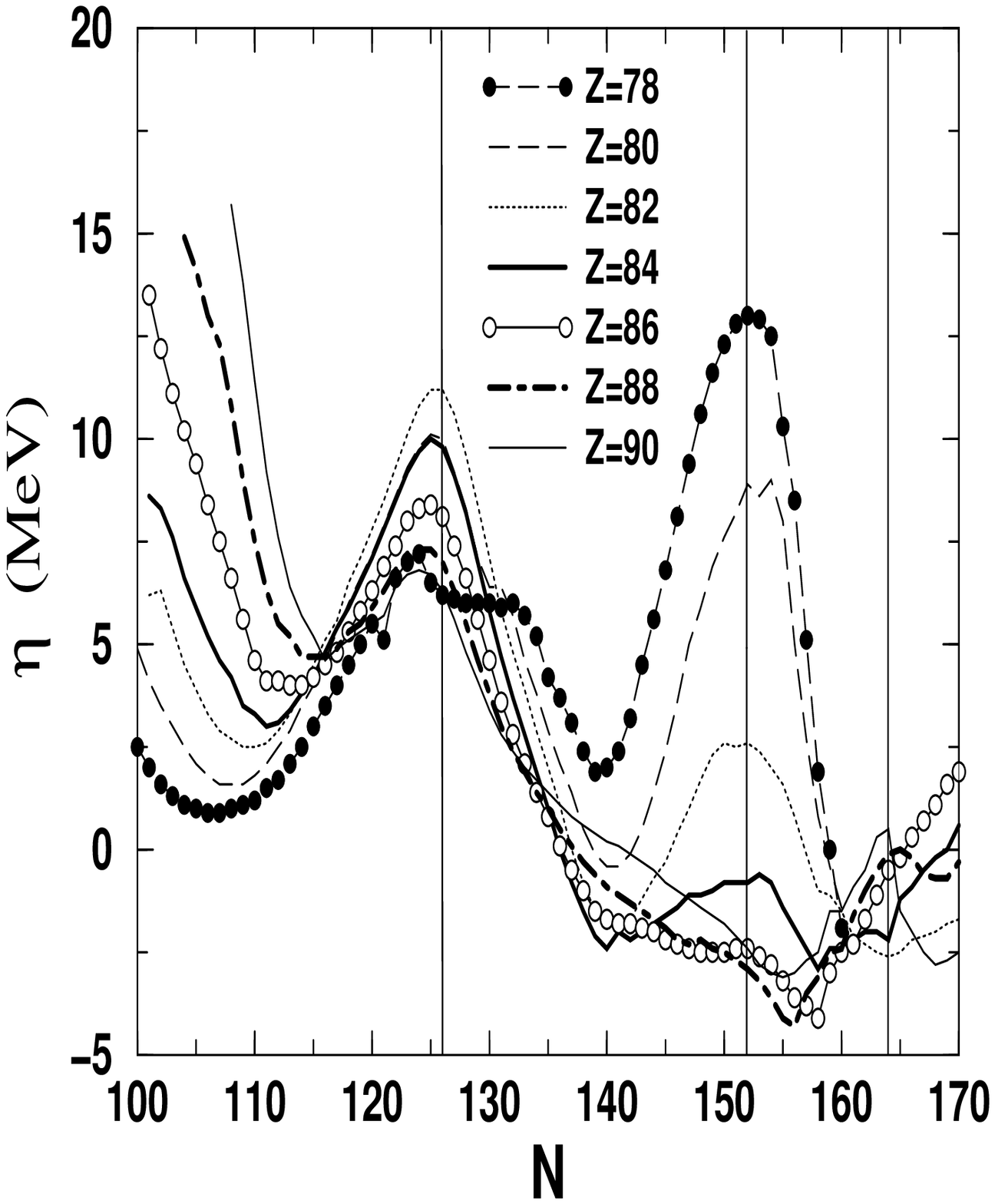}
\end{center}
\caption{Same as Fig 3 but for $\eta$ isolines.}
\label{4}
\end{figure}
\end{document}